\documentclass[11pt]{article}
\usepackage{jheppub}
\usepackage{bbm}
\usepackage{empheq}
\usepackage{bbm}
\usepackage[utf8]{inputenc}
\def\Om{{\mathcal{O}}}
\def\Cm{{\mathcal{C}}}
\def\Em{{\mathcal{E}}}
\def\Am{{\mathcal{A}}}

\def\Bm{{\mathcal{B}}}
\def\Nm{{\mathcal{N}}}

\def\Jm{{\mathcal{J}}}
\def\Im{{\mathcal{I}}}
\def\Km{{\mathcal{K}}}
\def\Qm{{\mathcal{Q}}}
\def\Lm{{\mathcal{L}}}
\def\Pm{{\mathcal{P}}}
\def\Dm{{\mathcal{D}}}

\def\Sm{{\mathcal{S}}}
\def\Mm{{\mathcal{M}}}

\def\a{\alpha}
\def\ad{\dot{\alpha}}
\def\b{\beta}
\def\bd{\dot{\beta}}

\def\e{\epsilon}
\def\m{\mu}
\def\md{\dot{\mu}}

\def\nd{\dot{\nu}}
\def\th{\theta}
\def\thb{\bar{\theta}}
\def\Th{\Theta}
\def\Thb{\bar{\Theta}}

\def\Xbfb{{\bar{\mathbf{X}}}}
\def\Xbf{{\mathbf{X}}}
\def\Zbf{{\mathbf{Z}}}

\def\pd{\partial}

\def\nn{\nonumber}

\title{\boldmath Mixed OPEs in $\Nm=2$ Superconformal Theories}

\author[a]{Israel A. Ram\'irez,}

\affiliation[a]{Departamento de F\'isica, Universidad T\'ecnica Federico Santa Mar\'ia, \\
Casilla 110-V, Valpara\'iso, Chile}

\emailAdd{ramirezkrause@gmail.com}

\abstract{
Using superspace techniques, we compute the mixed OPE between an ${\mathcal N}=2$ stress-tensor multiplet, a chiral multiplet and a flavor current multiplet. We perform a detailed analysis of the three-point function between two of the mentioned multiplets and a third arbitrary operator. We then solve all the constraints coming from the ${\mathcal N}=2$ superconformal symmetry and from the equations of motion and/or conservation equations, and obtain all the possible operators that can appear in the expansion. This calculation is the first step towards a more general superconformal block analysis of mixed correlators in ${\mathcal N}=2$ theories.%
}

\begin{document}
\maketitle
\flushbottom

\section{Introduction}

Lagrangian methods seem to be insufficient when studying $\Nm=2$ SCFTs. Although a large class of them are Lagrangian theories \cite{Bhardwaj:2013qia},  there are many strongly coupled fixed points which seem to not allow a Lagrangian description \cite{Gaiotto:2009we,Gaiotto:2009hg}. With the goal of developing a Lagrangian-free framework based only on the operator algebra, in \cite{Beem:2014zpa} the conformal bootstrap program for $\Nm = 2$ theories was initiated. The conformal bootstrap \cite{Ferrara:1973yt,Polyakov:1974gs,Mack:1975jr} has received renewed attention after the work of \cite{Rattazzi:2008pe}. The idea behind this approach is simple: imposing only unitarity and crossing symmetry for the four-point function, several CFT quantities can be obtained. 

As pointed out in \cite{Beem:2014zpa}, there are three classes of short representations which are directly related to physical characteristics of $\Nm=2$ theories, and thus can be regarded as a natural first step in the bootstrap program: the stress-tensor multiplet, the $\Nm=2$ chiral multiplets and the flavor current multiplet. By bootstrapping them, we expect to obtain relevant information about the $a$ and $c$ anomalies, the Coulomb branch, the Higgs branch and the flavor central charge $k$, among other relations. Following this election of multiplets, they studied the four-point function of chiral operators and the four-point function of flavor current multiplets, obtaining several numerical bounds. There was a technical reason why the stress-tensor four point function was not studied in \cite{Beem:2014zpa}: its conformal block expansion is not known. The block expansion for mixed operators is even more elusive.

Although the conformal block (or partial wave) decomposition of the four-point function is an essential ingredient in the conformal bootstrap program, there is no unified framework to compute the conformal blocks for different types of operators. Harmonic superspace techniques have proved useful to obtain the superconformal block expansion of $\frac{1}{2}$-BPS operators \cite{Dolan:2004mu,Bissi:2015qoa}, such as the flavor current multiplet. For the four-point function of two chiral and two anti-chiral operators, instead of harmonic superspace, chiral superspace has proven more useful \cite{Fitzpatrick:2014oza}. The stress-tensor multiplet is not $\frac{1}{2}$-BPS nor chiral, but rather ``semi-short" according to the classification of \cite{Dolan:2002zh}. A first step towards its block decomposition was taken in \cite{Liendo:2015ofa}, where, using standard Minkowski superspace techniques, the complete OPE of two stress-tensor multiplets was obtained. Due to the different nature of the three multiplets which we want to study in this article, $\Nm=2$ Minkowski superspace seems suitable when dealing with a mixed combination of them. We denote the corresponding operators of the stress-tensor, the chiral and the flavor current multiplets as $\Jm$, $\Phi$ and $\Lm_{i\,j}$, respectively.

%To achieve this, the existence of a $\Nm=2$ conserved current, $\Jm$, containing the stress-tensor \cite{Sohnius:1978pk} was used. Not only the stress-tensor multiplet has a $\Nm=2$ operator associated to it, both the chiral and conserved flavor current multiplets also have a $\Nm=2$ operator associated to them, $\Phi$ and $\Lm_{(i\,j)}$ respectively, and for the anti-chiral multiplet is known as $\bar\Phi$. Therefore, $\Nm=2$ superspace techniques seem very suitable when dealing with a mixed combination of them.

Another source of information used in \cite{Beem:2014zpa} was the existence of a protected subsector of operators, present in every $\Nm=2$ theory, that are isomorphic to a two-dimensional chiral algebra \cite{Beem:2013sza}. Using this correspondence between $\Nm=2$ theories and chiral algebras, along with the block decomposition of the flavor current four-point function, bounds involving the central charge $c$ and the flavor central charge $k$ were obtained \cite{Beem:2013sza}. Following the same spirit, and using the $\Jm\times\Jm$ OPE, bounds to the central $c$ were obtained \cite{Lemos:2015awa}. Furthermore, studying mixed correlators in the chiral algebra setup, yet another bound relating $c$ and $k$ was obtained \cite{Lemos:2015orc}. As pointed out in \cite{Lemos:2015orc}, an interesting result is obtained when combining the aforementioned analytical bounds involving both $c$ and $k$: all the canonical rank one SCFTs associated to maximal mass deformations of the Kodaira singularities with flavor symmetry $G=A_1,A_2,D_4,E_6,E_7,E_8$ \cite{Argyres:1995xn,Minahan:1996fg,Minahan:1996cj,Cheung:1997id} live at the intersection of the analytical bounds. It was also shown that the predicted theories with flavor symmetry $G=G_2,F_4$ \cite{Beem:2014zpa}, which have no known corresponding SCFT, live at the intersection of the bounds as well.

Those previous results entice us to keep studying systems of mixed correlators. While the single correlator bootstrap has already given interesting results, the addition of mixed correlators will give us access to the canonical rank one CFTs that live at the intersection of the analytical bounds. With the numerical bootstrap for the mixed system we will be able to explore CFT data inaccessible from the chiral algebra. Here we take a first step towards the construction of the superconformal blocks by obtaining the system of mixed OPE containing the three multiplets mentioned above: the stress-tensor multiplet, the chiral multiplets and the flavor current multiplet.

The outline of this article is as follows. In Section 2 we review $\Nm=2$ superconformal three-point function, presenting all the ingredients needed to solve the OPE. In Section 3, after introducing the EOMs and conservation equations of the $\Jm$, $\Phi$ and $\Lm_{i\,j}$ superfields we show how to solve,
\begin{align}
 \langle \Phi\,\Jm\,\Om \rangle\,,\qquad  \langle \Phi\,\Lm_{i\,j}\,\Om \rangle \,, \qquad  \langle \Jm\,\Lm_{i\,j}\,\Om \rangle\,,
\end{align} 
for every $\Om$ operator. This information allows us to write down the $\Phi \times \Jm$, $\Phi\times \Lm_{i\,j}$ and $\Jm\times\Lm_{i\,j}$ mixed OPEs. We end in Section 4 with conclusions. We also provide two appendices for notations and convention, plus solutions to the $\Om$ operators not listed in the OPEs.

%Even if bootstraping a system of mixed correlators has lead to tremendous success in the $O(N)$ models \cite{Kos:2014bka,Kos:2015mba}, our motivation

\section{The three-point function of $\Nm=2$ SCFT}

It is well known that conformal symmetry fixes, up to an overall constant, the two- and three-point function for any operator. For a recent review see \cite{Rychkov:2016iqz}. Superconformal symmetry also imposes restrictions to the form of the two- and three-point functions \cite{Osborn:1998qu,Park:1999pd}. The general expression for three-point functions in $\Nm=2$ superspace is,\footnote{We will follow the notation and conventions of \cite{Kuzenko:1999pi}, and we will also borrow some results from there.}
\begin{align}
 \langle \Om^{(1)}_{\Im_1}(z_1)\Om^{(2)}_{\Im_2}(z_2)\Om^{(3)}_{\Im_3}(z_3) \rangle=& \frac{T^{(1) ~\Jm_1}_{~\Im_1}\left(\hat u(z_{13}) \right)T^{(2) ~\Jm_2}_{~\Im_2}\left(\hat u(z_{23}) \right)}{(x_{\bar 1\,3})^{2\bar q_1}(x_{\bar 3\,1})^{2 q_1}(x_{\bar 2\,3})^{2\bar q_2}(x_{\bar 3\,2})^{2q_2}} H_{\Jm_1 \Jm_2 \Im_3}\left(\Zbf_3 \right)\,, \label{3ptf}
\end{align}
where $z^A=(x^a,\th_i^\a,\bar\th^{i\,\ad})$ is the supercoordinate, $q$ and $\bar q$ are given by $\Delta=q+\bar{q}$ and $r=q-\bar{q}$, $r$ being the $U(1)_r$-charge. The $\Im=(\a,\ad,R,r)$ is a collective index that labels all the irreducible representation to which $\Om$ belongs, it can also contain flavor indices. $H_{\Jm_1 \Jm_2 \Im_3}$ transforms as a tensor at $z_3$ in such a way that \eqref{3ptf} is covariant. The chiral and anti-chiral coordinates are,
\begin{align}
& x^{\ad \a}_{\bar{1}2}  = - x^{\ad \a}_{2\bar{1}} = x^{\ad \a}_{1-} -  x^{\ad \a}_{2+} -4\mathrm{i}\, \theta^{\a}_{2\,i} \bar{\theta}^{\ad i}_1\, ,
\\
& \theta_{12} = \theta_1 - \theta_2\, , \qquad  \bar{\theta}_{12} = \bar{\theta}_1 - \bar{\theta}_2\, ,
\end{align}
with $x^{\ad \a}_{\pm} = x^{\ad \a} \mp 2{\rm i}\th_{i}^\a \thb^{\ad\, i}$. The $\hat u$ matrices are defined as,
\begin{align}
 \hat u_i^{~j}(z_{12})=\left(\frac{{x_{\bar 2\, 1}}^2}{{x_{\bar 1\, 2}}^2} \right)^{1/2} \left(\delta_i^j-4{\rm i}\frac{\th_{12\,i}x_{\bar 1\,2}\bar\th_{12}^j}{{x_{\bar 1\,2}}^2} \right)\,. \label{hatu}
\end{align}
The argument of $H$ is given by three superconformally covariant coordinates $\mathbf{Z}_3 = (\Xbf_3, \Theta_3, \bar{\Theta}_3)$, which are defined as,
\begin{align}
&
\mathbf{X}_{3\,\a\, \ad}  = \frac{x_{3\bar{1}\,\a \bd}{ x_{\bar{1}2}^{\bd\b} }x_{2\bar{3}\,\b\ad}}{(x_{3\bar{1}})^2 (x_{2\bar{3}})^2}\, ,
&
&
\bar{\mathbf{X}}_{3\,\a\ad} = \mathbf{X}^{\dagger}_{3\,\a\ad}  
= -\frac{x_{3\bar{2}\, \a\bd} x_{\bar{2}1}^{\bd\b} x_{1\bar{3}\,\b\ad}}{(x_{3\bar{2}})^2  (x_{1\bar{3}})^2}\, ,
&
\\
&
\Theta^{i}_{3\, \a} = \mathrm{i} \left(\frac{x_{\bar{2}3\, \a \ad}}{x_{\bar{2}3}^2}\bar{\theta}^{\,\ad i}_{32} 
- \frac{x_{\bar{1}3\, \a \ad}}{x_{\bar{1}3}^2}\bar{\theta}^{\,\ad i}_{31} \right)\, ,
&
&
\bar{\Theta}_{3\, \ad\, i} = \mathrm{i} \left(\theta^{\a}_{32\, i} \frac{x_{\bar{3}2\, \a \ad}}{x_{\bar{3}2}^2} 
- \theta^{\a}_{31\, i} \frac{x_{\bar{3}1\, \a \ad}}{x_{\bar{3}1}^2} \right)\, \label{XXBTTb}.
&
\end{align}
An important relation which will play a key role in our computations is,
\begin{align}
 \bar{\mathbf{X}}_{3\,\a\ad} = \mathbf{X}_{3\,\a\ad}-4 \mathrm{i}\, \Theta^{i}_{3\,\a} \bar{\Theta}_{3\,\ad\, i}\, .
\end{align}
In addition, the function $H$ satisfies the scaling condition,
\begin{align}
H^{\Im}(\lambda \bar{\lambda}\Xbf_3,\lambda \Theta_3,\bar{\lambda} \bar{\Theta}_3) = \lambda^{2a} \bar{\lambda}^{2\bar{a}} H^{\Im}(\Xbf_3,\Theta_3,\bar{\Theta}_3)\, ,  \label{scaling} 
\end{align} 
with $a-2\bar{a} = 2-q$ and $\bar{a}-2a = 2-\bar{q}$. This last piece of information will help us identify the operator $\Om^{(3)}$ by comparing its quantum numbers with all the possible representations listed in Tab. \ref{repre}.

The conformally covariant operators $\Dm_{A}=\left(\partial /\partial \Xbf^a_3,\Dm_{\a\,i},\bar\Dm^{\ad\,i} \right)$ and $\Qm_{A}=\left(\partial /\partial \Xbf^a_3,\Qm_{\a\,i},\bar\Qm^{\ad\,i} \right)$, given by,
\begin{align}
\label{Ds}\begin{split}
\Dm_{\ad\, i} =  \frac{\pd}{\pd \Theta^{\a\, i}_3} + 4\mathrm{i} \bar{\Theta}^{\ad}_{3\,i} \frac{\pd}{\pd \Xbf_3^{\ad \a}}\, ,
\qquad
\bar{\Dm}^{\ad\, i} =  \frac{\pd}{\pd \bar{\Theta}_{3\, \ad\, i}} \, ,\\
\bar{\Qm}^{\ad\, i} =  \frac{\pd}{\pd \bar{\Theta}_{3\, \ad\, i}} -4\mathrm{i} \Theta^{i}_{3\,\a} \frac{\pd}{\pd \Xbf_{3\, \a \ad}}\, ,
\qquad
\Qm_{\a\, i}= \frac{\pd}{\pd \Theta^{\a\, i}_3}\, ,\end{split}
\end{align}
appear naturally when applying the superderivatives on $H\left( \Zbf\right)$:
\begin{align}\begin{split}
 D_{1\,\a}^{~i}H\left(\Xbf_3,\Th_3\,\Thb_3 \right)=&{\rm i} \frac{(x_{\bar 3\, 1})_{\a\,\bd}}{\left( {x_{\bar 1\, 3}}^2 {x_{\bar 3\, 1}}^2 \right)^{1/2}} \hat u_j^{~i}(z_{31})\bar{\Dm}^{\bd j} H\left( \Xbf_3,\Th_3\,\Thb_3 \right)\,, \\
 \bar D_{1\,\bd\,j}H\left(\Xbf_3,\Th_3\,\Thb_3 \right)=&{\rm i} \frac{(x_{\bar 1\, 3})_{\a\,\bd}}{\left( {x_{\bar 1\, 3}}^2 {x_{\bar 3\, 1}}^2 \right)^{1/2}} \hat u_j^{~i}(z_{13}){\Dm}^{\a}_i H\left( \Xbf_3,\Th_3\,\Thb_3 \right)\,, \\
 D_{2\,\a}^{~i}H\left(\Xbf_3,\Th_3\,\Thb_3 \right)=&{\rm i} \frac{(x_{\bar 3\, 2})_{\a\,\bd}}{\left( {x_{\bar 2\, 3}}^2 {x_{\bar 3\, 2}}^2 \right)^{1/2}} \hat u_j^{~i}(z_{32})\bar{\Qm}^{\bd j} H\left( \Xbf_3,\Th_3\,\Thb_3 \right)\,, \\
 \bar D_{2\,\bd\,j} H\left(\Xbf_3,\Th_3\,\Thb_3 \right)=&{\rm i} \frac{(x_{\bar 2\, 3})_{\a\,\bd}}{\left( {x_{\bar 2\, 3}}^2 {x_{\bar 3\, 2}}^2 \right)^{1/2}} \hat u_j^{~i}(z_{23}){\Qm}^{\a}_i H\left( \Xbf_3,\Th_3\,\Thb_3 \right)\,.\end{split} \label{dd}
\end{align}

There are similar relations for quadratic derivatives. A quick computation shows,
\begin{align}
 D_1^{~\a\,i}D_{1\,\a}^{~j} H\left(\Xbf_3,\Th_3\,\Thb_3 \right)=-\frac{\hat u_{k}^{~i}(z_{1\,3})\hat u_{l}^{~j}(z_{1\,3})}{{x_{\bar 1\,3}}^2 {x_{\bar 3\, 1}}^2} \bar \Dm_{\ad}^k\bar \Dm^{\ad \,l}H\left(\Xbf_3,\Th_3\,\Thb_3 \right)\,,
\end{align}
and similar relations for $\bar D_{1\,i\,j}$, $D_2^{~i\,j}$ and $\bar D_{2\,i\,j}$. These relations will be very important when we impose the EOM/conservation equations of the superfields on the three-point function, restricting the form of all possible $\Om$ in \eqref{3ptf}.

For a general CFT, the conformal symmetry is strong enough
to to fix the OPE coefficients of the descendants in terms of that of the primary operator. This is not the
case in supersymmetric theories, where nilpotent structures which
contribute to the superdescendants can appear in the three-point
function, see for example equation (3.23) in \cite{Kuzenko:1999pi}, and also 
equations (3.18) and (3.25) in \cite{Liendo:2015ofa}. In the cases studied here, the EOM/conservation
equations will impose restrictions strong enough to fix the form of
the three-point function completely, but this is not necessarily true
for general operators.

%%%%%%%%%%%%%%%%%%%%%%%%%%%%%%%%%%%%%%%%%%%%%%%%%%%%%%%%%%
%%%%%%%%%%%%%%%%%%%%%%%%%%%%%%%%%%%%%%%%%%%%%%%%%%%%%%%%%%
%%%%%%%%%%%%%%%%%%%%%%%%%%%%%%%%%%%%%%%%%%%%%%%%%%%%%%%%%%

\section{Mixed OPE}

We mentioned in the introduction that we are interested in the mixed OPEs of three multiplets: the stress-tensor multiplet, the $\Nm=2$ chiral multiplets and the flavor current multiplet, because of their close relation with physical properties of $\Nm=2$ theories:
 
$\bullet$ The semi-short multiplet $\hat \Cm_{0(0,0)}$\footnote{We will mostly follow the conventions of \cite{Dolan:2002zh}, see also Tab. \ref{repre} for a summary. 
} contains a conserved current of spin 2 and the spin 1 R-symmetry currents. It is well known that such spin 2 conserved current is the stress-tensor, which is present in every local theory, therefore, the study of this multiplet will give us general information about $\Nm=2$ theories. Its higher spin generalization $\hat\Cm_{0\left(j_1,j_2 \right)}$ will contain higher spin conserved currents, which are not expected to appear in interacting theories \cite{Maldacena:2011jn,Alba:2013yda}.

$\bullet$ The vacuum expectation value of chiral multiplets, $\Em_q$,\footnote{We define $\Em_q:=\Em_{q(0,0)}$. Although chiral operator with higher spin, $\Em_{q(j,0)}$ are allowed by representation theory, see Tab. \ref{repre}, it was shown that such multiplets are absent in every known $\Nm=2$ theory \cite{Buican:2014qla}.} parametrizes the Coulomb branch of the moduli space of $\Nm=2$ theories. The complex dimension of this branch defines the rank of the $\Nm=2$ theory. For a systematic study of rank one theories using their Coulomb branch geometries see \cite{Argyres:2015ffa,Argyres:2015gha}. 

$\bullet$ The $\hat \Bm_1$ multiplet plays an important role in theories with flavor symmetry. Global symmetries currents can only appear in the $\hat \Bm_1$ or the $\hat \Cm_{0\left(\frac{1}{2},\frac{1}{2} \right)}$ multiplets. We already argued why this last multiplet must be absent. Therefore, for the study of $\Nm=2$ theories with flavor symmetries the $\hat\Bm_1$ multiplet is fundamental. In analogy with the relation between chiral multiplets and the Coulomb branch, information about the Higgs branch is encoded in the $\hat \Bm_R$ multiplets. 

As already noted, all of these multiplets are described by an $\Nm=2$ superfield with a well known EOM/conservation equation. Indeed, the $\Nm=2$ superspace conserved current associated to the stress-tensor, which we denote as $\Jm$, satisfies the conservation equations \cite{Sohnius:1978pk},
\begin{subequations}
\label{eomJ}
\begin{align}
 D^{i\,j}\Jm(z)=&0 \,,\label{eomJ1} \\
 \bar D^{i\,j}\Jm(z)=&0 \,.\label{eomJ2} 
\end{align}
\end{subequations}
The chiral multiplets $\Em_q$ are described by an $\Nm=2$ chiral superfield, denoted here by $\Phi$, satisfying a linear equation,
\begin{align}
 D^{\ad\,i}\Phi(z)=0\,, \label{eomP}
\end{align}
which is the same for every $r$-charge. Unitarity implies $q\ge1$. Because $\Em_1$ is free, we will only consider the $q>1$ cases. Finally, just as with the stress-tensor multiplet, the $\Nm=2$ flavor current superfield, which we call $\Lm_{(i\,j)}$, satisfies the conservation equations,
\begin{subequations}
\label{eomL}
\begin{align}
D^\a_{(i}\Lm_{j\,k)}(z)=&0 \,,\label{eomL1}\\
\bar D^{\ad} _{(i}\Lm_{j\,k)}(z)=&0 \,. \label{eomL2}
\end{align}
\end{subequations}

Below, we solve the three-point function in order to obtain the OPE. We will first solve the OPE $\Em_q\times\hat\Cm_{0(0,0)}$. The reason is twofold: first, it has been shown that a chiral field imposes a very strong constraint to the form of the three-point function, see for example \cite{Poland:2010wg,Beem:2014zpa,Lemos:2015awa}; second, since $\Jm$ carries no indices, possible solutions to the three-point function are, a priori, simpler than solutions with $\Lm_{i\,j}$. The solutions of $H\left(\Zbf \right)^\Im$ tell us the quantum numbers of $\Om^\Im$. Knowledge of the quantum numbers allows us to identify the $\Om^\Im$ multiplet with the corresponding long, short or semi-short multiplet, see Tab. \ref{repre}. Following this logic we next solve the OPE $\Em_q\times\hat\Bm_1$. We end this section with the $\hat\Cm_{0(0,0)}\times\hat\Bm_1$ OPE. 
%%%%%%%%%%%%%%%%%%%%%%%%%%%%%%%%%%%%%%%%%%%%%%%%%%%%%%%%%%%%%%%%%
%%%%%%%%%%%%%%%%%%%%%%%%%%%%%%%%%%%%%%%%%%%%%%%%%%%%%%%%%%%%%%%%%

%%%%%%%%%%%%%%%%%%%%%%%%%%%%%%%%%%%%%%%%%%%%%%%%%%%%%%%%%%%%%%%%%
%%%%%%%%%%%%%%%%%%%%%%%%%%%%%%%%%%%%%%%%%%%%%%%%%%%%%%%%%%%%%%%%%

\subsection{$\Em_q\times\hat \Cm_{0(0,0)}$}

The three-point function \eqref{3ptf} for a chiral operator and the stress tensor multiplet is, 
\begin{align}
 \langle \Phi(z_1)\, \Jm (z_2)\, \Om^I(z_3) \rangle=\frac{\lambda_{\Phi\Jm\Om}}{(x_{\bar 3\, 1}^2)^q x_{\bar 2\, 3}^2 x_{2\,\bar 3}^2 }H^I(\Zbf_3)\,. \label{pjo}
\end{align}

The EOM of $\Phi$ \eqref{eomP} and the conservation equation of $\Jm$ \eqref{eomJ} will imply restrictions in the form of conformally covariant operators acting on $H\left(\Zbf \right)$,
\begin{align}
\bar D^{\ad\, i}_1 \langle \Phi(z_1)\, \Jm (z_2)\, \Om^I(z_3) \rangle=&0 \qquad \Rightarrow\qquad \Dm_{\a\, j}\, H^I(\Zbf_3)=0\,, \label{pjo:c1}\\
D^{ij}_2 \langle \Phi(z_1)\, \Jm (z_2)\, \Om^I(z_3) \rangle=&0 \qquad \Rightarrow\qquad  \bar\Qm^k_{\ad}\, \bar\Qm^{\ad\, l}\,H^I(\Zbf_3)=0\, , \label{pjo:c2}\\
\bar D^{ij}_2 \langle \Phi(z_1)\, \Jm (z_2)\, \Om^I(z_3) \rangle=&0 \qquad \Rightarrow\qquad \Qm^k_\a\, \Qm^{\a\, l}\,H^I(\Zbf_3)=0\, ,\label{pjo:c3}
\end{align}
see  \eqref{dd}. The $\Dm$ and $\Qm$ operators were defined in \eqref{Ds}.

The first constraint, \eqref{pjo:c1}, implies $H(\Xbf,\Theta,\Bar \Theta)=H(\Xbf+2\,\mathrm{i}\,\Theta\,\sigma\,\bar\Theta\,,\bar\Theta)=H(\bar\Xbf\,,\bar\Theta)$ (we omit the subscript 3 from now on.) Since $\bar\Qm\, \Xbfb=0$, \eqref{pjo:c2} implies that $H\left(\Xbfb,\Thb \right)$ can have at most a quadratic term in $\Thb$ in the form of $\Thb^{\ad}_i\,\Thb^{\bd\,i}=\Thb^{\ad\,\bd}$ \cite{Kuzenko:1999pi}. Thus our solutions are of the form $H(\bar \Xbf,\bar\Theta)=f(\bar \Xbf)+g(\bar\Xbf)_{\ad\, i}\,\bar\Theta^{\ad\, i}+h(\bar\Xbf)_{\ad\,\bd}\,\bar\Theta^{\ad\,\bd}$. At this point is good to note that it is not possible, using only $\bar\Xbf$, to construct functions $f$, $g$ and $h$ with any $SU(2)_R$- nor $U(1)_r$-charges. Therefore, we can, and will, study the solutions of the $f$, $g$ and $h$ terms separately. The constraint \eqref{pjo:c3} implies,
\begin{align}
\label{Hpjo:c2}
 \frac{\partial^2}{\partial \Theta^{\a\, i}\partial\Theta^j_{\a}}H(\Zbf)=-4\left(\bar\Box f(\bar\Xbf)+\bar\Box g(\bar\Xbf)_{\ad\, k}\,\bar\Theta^{\ad\, k}+\bar\Box h(\bar\Xbf)_{\ad\,\bd}\,\bar\Theta^{\ad\,\bd} \right)\bar\Theta^{\md}_i\,\bar\Theta_{\md\, j}=0\, ,
\end{align}
where we defined $\bar\Box=\frac{\partial^2}{\partial\bar\Xbf^a\partial\bar\Xbf_a}$. A quick computation shows that $\bar\Theta^{\md}_i\bar\Theta_{\md\, j}\bar\Theta^{\ad\,\bd}$ is always vanishing. This will generate solutions to \eqref{pjo} with arbitrary conformal dimension. We will identify those solutions with long supermultiplets. Furthermore, $\bar\Theta^{\md}_i\,\bar\Theta_{\md\, j}\,\bar\Theta^{\ad\, k}$ does not impose any new condition, thus, the solutions to $\bar\Box f\left(\Xbfb \right)=0$ are also solutions to $\bar\Box g\left(\Xbfb \right)=0$. The physical solutions to \eqref{Hpjo:c2} are,
\begin{subequations}\label{ExC:s}
\begin{align}
 \text{Multiplet} \qquad & H\left(\Zbf \right) \nn\\
 \Am^\Delta_{0,3-q\left(\frac{\ell}{2},\frac{\ell+2}{2} \right)}: \qquad & \left( \Xbfb^2\right)^{-\frac{3}{2}+\frac{\Delta-\ell-q}{2}}\,\Xbfb_{\a_1\,\ad_1}\cdots \Xbfb_{\a_\ell\,\ad_\ell}\Thb_{\ad_{\ell+1}\,\ad_{\ell+2}}\,, \label{ExC:l1}\\
 \Am^\Delta_{0,3-q,\left(\frac{\ell}{2},\frac{\ell}{2} \right)}: \qquad & \left( \Xbfb^2\right)^{-\frac{3}{2}+\frac{\Delta-\ell-q}{2}}\Xbfb_{\a_1\,\ad_1}\cdots \Xbfb_{\a_{\ell-1}\,,\,\ad_{\ell-1}}\Xbfb_{\a_{\ell}\, \md}\epsilon^{\md\,\bd}\Thb_{\ad_{\ell}\,\bd} \,,\label{ExC:l2}\\
 \Am^\Delta_{0,3-q,\left(\frac{\ell+2}{2},\frac{\ell}{2} \right)}: \qquad & \left( \Xbfb^2\right)^{-\frac{5}{2}+\frac{\Delta-\ell-q}{2}}\Xbfb_{\a_1\,,\,\ad_1}\cdots \Xbfb_{\a_\ell\,,\,\ad_\ell}\Xbfb_{\a_{\ell+1}\, \md}\Xbfb_{\a_{\ell+2}\, \nd}\Thb^{\md\,\nd}\,,\label{ExC:l3}\\
 \bar \Cm_{0,-q\left(\frac{\ell}{2},\frac{\ell}{2} \right)}: \qquad & \Xbfb_{\a_1\,\ad_1}\cdots \Xbfb_{\a_{\ell}\,\ad_{\ell}} \,,\\
 \bar \Cm_{\frac{1}{2},\frac{3}{2}-q\left(\frac{\ell}{2},\frac{\ell+1}{2} \right)}: \qquad & \Xbfb_{\a_1\,\ad_1}\cdots \Xbfb_{\a_{\ell}\,\ad_{\ell}}\Thb^i_{\ad_{\ell+1}} \,,\\
 \Cm_{\frac{1}{2},-\frac{1}{2}\left(0,\frac{1}{2} \right)}: \qquad & \left(\Xbfb^2 \right)^{-1}\Thb^i_{\ad}\,, \\
 \bar\Bm_{\frac{1}{2},\frac{3}{2}-q\left(\frac{1}{2},0\right)}: \qquad & \left(\Xbfb^2 \right)^{-2}\Xbfb_{\ad\,\md}\Thb^{\md\,i}\,,\\
 \bar \Em_{-q(0,0)}: \qquad & \left(\Xbfb \right)^{-1} \,. \label{ExC:e}
\end{align}
\end{subequations}
There are also extra solutions to \eqref{pjo} which we have discarded, see \eqref{ExC:nu} and \eqref{ExC:n4}.

When a long multiplet hits its unitarity bound, it decomposes into different (semi-)short multiplets \cite{Dolan:2002zh}. The unitarity bounds of our three long multiplets \eqref{ExC:l1}, \eqref{ExC:l2} and \eqref{ExC:l3} depend on the value of $q$. There are three distinctive ranges in every case. For \eqref{ExC:l1} its decomposition is,
\begin{align}\label{ExC:lub1}\begin{split}
 q<2 :\qquad  &\Am^{5-q+\ell}_{0,3-q\left(\frac{\ell}{2},\frac{\ell+2}{2} \right)}\sim \Cm_{0,3-q\left(\frac{\ell}{2},\frac{\ell+2}{2} \right)}+\Cm_{\frac{1}{2},\frac{7}{2}-q\left(\frac{\ell-1}{2},\frac{\ell+2}{2} \right)}\,,\\
       q=2 :\qquad  &\Am^{3+\ell}_{0,1\left(\frac{\ell}{2},\frac{\ell+2}{2} \right)}\sim \hat \Cm_{0\left(\frac{\ell}{2},\frac{\ell+2}{2} \right)}+\hat \Cm_{\frac{1}{2}\left(\frac{\ell-1}{2},\frac{\ell+2}{2} \right)}+\hat \Cm_{\frac{1}{2}\left(\frac{\ell}{2},\frac{\ell+1}{2} \right)}+\hat\Cm_{1\left(\frac{\ell-1}{2},\frac{\ell+1}{2} \right)}\,,\\
        q>2 :\qquad  &\Am^{1+q+\ell}_{0,3-q\left(\frac{\ell}{2},\frac{\ell+2}{2} \right)}\sim \bar\Cm_{0,3-q\left(\frac{\ell}{2},\frac{\ell+2}{2} \right)}+\bar\Cm_{\frac{1}{2},\frac{7}{2}-q\left(\frac{\ell}{2},\frac{\ell+1}{2} \right)}\,.
\end{split}\end{align}
For \eqref{ExC:l2} the decomposition is,
\begin{align}\label{ExC:lub2}\begin{split}
        q<3 :\qquad&  \Am^{5-q+\ell}_{0,3-q\left(\frac{\ell}{2},\frac{\ell}{2} \right)}\sim \Cm_{0,3-q\left(\frac{\ell}{2},\frac{\ell}{2} \right)}+\Cm_{\frac{1}{2},\frac{7}{2}-q\left(\frac{\ell-1}{2},\frac{\ell}{2} \right)}\,,\\
        q=3 :\qquad&  \Am^{2+\ell}_{0,0\left(\frac{\ell}{2},\frac{\ell}{2} \right)}\sim \hat \Cm_{0\left(\frac{\ell}{2},\frac{\ell}{2} \right)}+\hat \Cm_{\frac{1}{2}\left(\frac{\ell-1}{2},\frac{\ell}{2} \right)}+\hat \Cm_{\frac{1}{2}\left(\frac{\ell}{2},\frac{\ell-1}{2} \right)}+\hat\Cm_{1\left(\frac{\ell-1}{2},\frac{\ell-1}{2} \right)}\,,\\
        q>3 :\qquad&  \Am^{-1+q+\ell}_{0,3-q\left(\frac{\ell}{2},\frac{\ell}{2} \right)}\sim \bar\Cm_{0,3-q\left(\frac{\ell}{2},\frac{\ell}{2} \right)}+\bar\Cm_{\frac{1}{2},\frac{7}{2}-q\left(\frac{\ell}{2},\frac{\ell-1}{2} \right)}\,.
\end{split}\end{align}
Finally, \eqref{ExC:l3} decomposes as,
\begin{align}\label{ExC:lub3}\begin{split}
         q<4 :\qquad&  \Am^{7-q+\ell}_{0,3-q\left(\frac{\ell+2}{2},\frac{\ell}{2} \right)}\sim \Cm_{0,3-q\left(\frac{\ell+2}{2},\frac{\ell}{2} \right)}+\Cm_{\frac{1}{2},\frac{7}{2}-q\left(\frac{\ell+1}{2},\frac{\ell}{2} \right)}\,,\\
         q=4 :\qquad&  \Am^{3+\ell}_{0,-1\left(\frac{\ell+2}{2},\frac{\ell}{2} \right)}\sim \hat \Cm_{0\left(\frac{\ell+2}{2},\frac{\ell}{2} \right)}+\hat \Cm_{\frac{1}{2}\left(\frac{\ell+1}{2},\frac{\ell}{2} \right)}+\hat \Cm_{\frac{1}{2}\left(\frac{\ell+2}{2},\frac{\ell-1}{2} \right)}+\hat\Cm_{1\left(\frac{\ell+1}{2},\frac{\ell-1}{2} \right)}\,,\\
         q>4 :\qquad&  \Am^{-1+q+\ell}_{0,3-q\left(\frac{\ell+2}{2},\frac{\ell}{2} \right)}\sim \bar\Cm_{0,3-q\left(\frac{\ell+2}{2},\frac{\ell}{2} \right)}+\bar\Cm_{\frac{1}{2},\frac{7}{2}-q\left(\frac{\ell+2}{2},\frac{\ell-1}{2} \right)}\,.
\end{split}\end{align}
Since our selection rules do not give any of the terms in the decomposition of the longs, we will follow the same procedure as in \cite{Liendo:2015ofa} and only take the first term of each decomposition in the OPE. The reason is simple: let us take, for example, the \eqref{ExC:l1} when $q<2$. As we can see in \eqref{ExC:lub1}, it decomposes into two semi-short multiplet: $\Cm_{0,3-q\left(\frac{\ell}{2},\frac{\ell+2}{2} \right)}$ and $\Cm_{\frac{1}{2},\frac{7}{2}-q\left(\frac{\ell-1}{2},\frac{\ell+2}{2} \right)}$. When we solve \eqref{pjo} imposing all the constraints \eqref{pjo:c1}, \eqref{pjo:c2} and \eqref{pjo:c3} we do not obtain any solution with the quantum numbers of $\Cm_{\frac{1}{2},\frac{7}{2}-q\left(\frac{\ell-1}{2},\frac{\ell+2}{2} \right)}$, therefore, our selection rules do not allow such multiplet as a solution to \eqref{pjo}. The other multiplet in the expansion, $\Cm_{0,3-q\left(\frac{\ell}{2},\frac{\ell+2}{2} \right)}$, is nothing but a special limit of \eqref{ExC:l1}, which is allowed by our selection rules. The reader might wonder if the selection rules ever allow a term in the decomposition of the long multiplet besides the first term, or maybe we are omitting valid solutions. Later, when studying the $\hat \Cm_{0(0,0)}\times \hat \Bm_1$ case, we will obtain a solution, \eqref{CxB:c5} and \eqref{CxB:c6}, which appear in the decomposition of a long multiplet \eqref{CxB:l1}.

Finally, we list the OPE between an $\Nm=2$ chiral and an $\Nm=2$ stress-tensor multiplet,
\begin{align}\begin{split}
 \Em_{q(0,0)}\times \hat\Cm_{0(0,0)}\sim & \Em_{q}+\Cm_{0,q\left(\frac{\ell}{2},\frac{\ell}{2} \right)} + \Cm_{\frac{1}{2},q-\frac{3}{2}\left(\frac{\ell+1}{2},\frac{\ell}{2} \right)}+ \Bm_{\frac{1}{2},q-\frac{3}{2}\left(0,\frac{1}{2}\right)} + \Am^\Delta_{0,q-3\left(\frac{\ell}{2},\frac{\ell+2}{2} \right)} \\
                                         & + \Am^\Delta_{0,q-3\left(\frac{\ell+1}{2},\frac{\ell+1}{2} \right)} + \Am^\Delta_{0,q-3\left(\frac{\ell+2}{2},\frac{\ell}{2} \right)} + {\mathcal F}_q\,, \end{split}\label{ExC:OPE}
\end{align}
where $\mathcal F_q$ is,
\begin{align}
 {\mathcal F}_q=\left\lbrace \begin{array}{lc}  \bar \Cm_{0,q-3\left(\frac{\ell}{2},\frac{\ell+2}{2} \right)} +\bar \Cm_{0,q-3\left(\frac{\ell+1}{2},\frac{\ell+1}{2} \right)} + \bar \Cm_{0,q-3\left(\frac{\ell+2}{2},\frac{\ell}{2} \right)} & q<2 \\ \bar \Cm_{0,-1\left(\frac{\ell}{2},\frac{\ell+2}{2} \right)} +\bar \Cm_{0,-1\left(\frac{\ell+1}{2},\frac{\ell+1}{2} \right)} + \hat \Cm_{0\left(\frac{\ell+2}{2},\frac{\ell}{2} \right)} + \bar \Cm_{\frac{1}{2},\frac{1}{2}\left( \frac{1}{2},0 \right)}& q=2 \\ \bar \Cm_{0,q-3\left(\frac{\ell}{2},\frac{\ell+2}{2} \right)} +\bar \Cm_{0,q-3\left(\frac{\ell+1}{2},\frac{\ell+1}{2} \right)} + \Cm_{0,q-3\left(\frac{\ell+2}{2},\frac{\ell}{2} \right)} & 3>q>2 \\ \bar \Cm_{0,0\left(\frac{\ell}{2},\frac{\ell+2}{2} \right)} +\hat \Cm_{0\left(\frac{\ell+1}{2},\frac{\ell+1}{2} \right)} + \Cm_{0,0\left(\frac{\ell+2}{2},\frac{\ell}{2} \right)} & q=3 \\\bar \Cm_{0,q-3\left(\frac{\ell}{2},\frac{\ell+2}{2} \right)} +\Cm_{0,q-3\left(\frac{\ell+1}{2},\frac{\ell+1}{2} \right)} + \Cm_{0,q-3\left(\frac{\ell+2}{2},\frac{\ell}{2} \right)} & 4>q>3 \\ \hat \Cm_{0\left(\frac{\ell}{2},\frac{\ell+2}{2} \right)} +\Cm_{0,1\left(\frac{\ell+1}{2},\frac{\ell+1}{2} \right)} + \Cm_{0,1\left(\frac{\ell+2}{2},\frac{\ell}{2} \right)} & q=4 \\ \Cm_{0,q-3\left(\frac{\ell}{2},\frac{\ell+2}{2} \right)} + \Cm_{0,q-3\left(\frac{\ell+1}{2},\frac{\ell+1}{2} \right)} + \Cm_{0,q-3\left(\frac{\ell+2}{2},\frac{\ell}{2} \right)} & q>4 \end{array} \right.\,,
\end{align}
and $\ell\ge0$.

%%%%%%%%%%%%%%%%%%%%%%%%%%%%%%%%%%%%%%%%%%%%%%
%%%%%%%%%%%%%%%%%%%%%%%%%%%%%%%%%%%%%%%%%%%%%%

%%%%%%%%%%%%%%%%%%%%%%%%%%%%%%%%%%%%%%%%%%%%%%
%%%%%%%%%%%%%%%%%%%%%%%%%%%%%%%%%%%%%%%%%%%%%%

\subsection{$\Em_{q}\times\hat \Bm_1$}

The three-point function \eqref{3ptf} whith $\Om^{(1)}=\Phi$ and $\Om^{(2)}=\Lm_{ij}$ is given by, 
\begin{align}
 \langle \Phi(z_1)\, \Lm_{ij} (z_2)\, \Om^I(z_3) \rangle=\lambda_{\Phi\Lm\Om} \frac{{\hat u}_{i}^{\; k}(z_{23}){\hat u}_{j}^{\; l}(z_{23})}{(x_{\bar 3\, 1}^2)^q x_{\bar 2\, 3}^2 x_{2\,\bar 3}^2 }H_{kl}^I(\Zbf)\,, \label{plo}
\end{align}
where the $\hat{u}$ matrices were defined in \eqref{hatu}. The symmetry $\Lm_{i\,j}=\Lm_{j\,i}$ must also appear in $H\left(\Zbf \right)$, implying $H_{mn}^\Im=H_{nm}^\Im$.

Just as with $\Jm$, the conservation equations for $\Lm_{ij}$ \eqref{eomL} imply constraints to $H\left(\Zbf \right)$,
\begin{subequations}
\label{Lcns}
\begin{align}
D^\a_{(i}\,\Lm(z)_{j\,k)}=&0 \qquad \Rightarrow \qquad \bar{\Qm}_{\ad\,(i}\,H_{mn)}=0 \,,\label{Lcns1}\\
\bar D^{\ad} _{(i}\,\Lm_{j\,k)}(z)=&0 \qquad \Rightarrow \qquad \Qm_{\a\,(i}\,H_{mn)}=0\,. \label{Lcns2}
\end{align}
\end{subequations}
Beside these conditions, we have the one that comes from the chiral supermultiplet \eqref{eomP}, but we already know from \eqref{pjo:c1} that it implies $H(\Zbf)_{mn}^\Im=H(\Xbfb,\Thb)_{mn}^\Im$.

Since $\bar \Qm_{\ad\,i}\,\Xbfb_{\m\,\md}=0$, we only need to expand $H(\Xbfb,\Thb)$ in powers of $\Thb$ and find which $\Thb$ structures satisfy \eqref{Lcns1}. There are only three of those structures,
\begin{align}
\epsilon^{(m|(i}\epsilon^{j)|n)}\,, \qquad \Thb^{\ad (i}\e^{j)m}\,, \qquad \Thb^{(i\,j)}\,. \label{ExB:thetasol}
\end{align}
Finally, we use $\Xbfb$ to construct all possible solutions to \eqref{Lcns2}. The solutions to \eqref{plo} are,
\begin{subequations}\label{ExB:sol}
\begin{align}
 \text{Multiplet} \qquad& H\left(\Zbf \right)\nn\\
 \Am^\Delta_{0,3-q\left(\frac{\ell}{2},\frac{\ell}{2} \right)}: \qquad &\left(\Xbfb^2 \right)^{-\frac{3}{2}+\frac{\Delta-q-\ell}{2}}\Xbfb_{\a_1\,\ad_1}\cdots \Xbfb_{\a_\ell\,\ad_\ell}\Thb^{i\,j}\,,\label{ExB:l1}\\
 \bar\Bm_{1,-q(0,0)}: \qquad &e^{(m|(i}\e^{j)|n)}\,,\\
 \bar\Cm_{\frac{1}{2},\frac{3}{2}-q\left(\frac{\ell}{2},\frac{\ell+1}{2} \right)}: \qquad &\Xbfb_{\a_1\,\ad_1}\cdots \Xbfb_{\a_\ell\,\ad_\ell}\Thb_{\ad_{\ell+1}}^{(i}\e^{j)m}\,.
\end{align}
\end{subequations}
For the only unphysical solution to \eqref{plo} see \eqref{ExB:nu}

The unitarity bound of our long multiplet \eqref{ExB:l1} depends on the $U(1)_r$-charge, in a similar fashion to the $\Em_q\times\hat\Cm_{0(0,0)}$ case. Its decomposition is,
\begin{align}\begin{split}
 q<3:\qquad &\Am^{5-q+\ell}_{0,3-q\left(\frac{\ell}{2},\frac{\ell}{2} \right)} \sim  \Cm_{0,3-q\left(\frac{\ell}{2},\frac{\ell}{2} \right)} + \Cm_{\frac{1}{2},\frac{7}{2}-q\left(\frac{\ell-1}{2},\frac{\ell}{2} \right)}\,,\\
 q=3:\qquad&\Am^{2+\ell}_{0,0\left(\frac{\ell}{2},\frac{\ell}{2} \right)} \sim  \hat \Cm_{0\left(\frac{\ell}{2},\frac{\ell}{2} \right)} + \hat \Cm_{\frac{1}{2}\left(\frac{\ell-1}{2},\frac{\ell}{2} \right)} + \hat\Cm_{\frac{1}{2}\left(\frac{\ell}{2},\frac{\ell-1}{2} \right)} + \hat \Cm_{1\left(\frac{\ell-1}{2},\frac{\ell-1}{2} \right)}\,,\\
 q>3:\qquad&\Am^{-1+q+\ell}_{0,3-q\left(\frac{\ell}{2},\frac{\ell}{2} \right)} \sim  \bar \Cm_{0,3-q\left(\frac{\ell}{2},\frac{\ell}{2} \right)} + \bar \Cm_{\frac{1}{2},\frac{5}{2}-q\left(\frac{\ell}{2},\frac{\ell-1}{2} \right)}\,.\end{split} \label{ExB:l}
\end{align}
Among the solutions that we found, there is no $\Cm_{\frac{1}{2},\frac{7}{2}-q\left(\frac{\ell-1}{2},\frac{\ell}{2} \right)}$, $\bar \Cm_{\frac{1}{2},\frac{5}{2}-q\left(\frac{\ell}{2},\frac{\ell-1}{2} \right)}$, $\hat \Cm_{\frac{1}{2}\left(\frac{\ell-1}{2},\frac{\ell}{2} \right)}$, $\hat \Cm_{\frac{1}{2}\left(\frac{\ell}{2},\frac{\ell-1}{2} \right)}$ nor $\hat \Cm_{1\left(\frac{\ell-1}{2},\frac{\ell-1}{2} \right)}$, therefore, we do not take them into account in the OPE, as explained before.

Finally, we list the OPE of $\Em_{q(0,0)}\times \hat\Bm_1$,
\begin{align}\label{ExB:OPE}
 \Em_q\times\hat\Bm_{1}\sim & \Bm_{1,q(0,0)}+ \Cm_{\frac{1}{2},q-\frac{3}{2}\left(\frac{\ell+1}{2},\frac{\ell}{2} \right)}+\Am^\Delta_{0,q-3\left( \frac{\ell}{2},\frac{\ell}{2} \right)}+\left\lbrace\begin{array}{lr} \bar \Cm_{0,q-3\left(\frac{\ell}{2},\frac{\ell}{2} \right)} & q<3\,, \\ \hat \Cm_{0\left(\frac{\ell}{2},\frac{\ell}{2} \right)} & q=3\,,\\ \Cm_{0,q-3\left(\frac{\ell}{2},\frac{\ell}{2} \right)} & q>3\,. \end{array} \right.
\end{align}
With $\ell\ge0$. Just as with the $\Em_q\times\hat\Cm_{0(0,0)}$ OPE, the $\Em_{q(0,0)}\times \hat\Bm_1$ OPE has a dependency on the value of $q$.

%%%%%%%%%%%%%%%%%%%%%%%%%%%%%%%%%%%%%%%%%%%%%%%%%%%%%%%%%%%%%%%%%%%%%%%
%%%%%%%%%%%%%%%%%%%%%%%%%%%%%%%%%%%%%%%%%%%%%%%%%%%%%%%%%%%%%%%%%%%%%%%

%%%%%%%%%%%%%%%%%%%%%%%%%%%%%%%%%%%%%%%%%%%%%%%%%%%%%%%%%%%%%%%%%%%%%%%
%%%%%%%%%%%%%%%%%%%%%%%%%%%%%%%%%%%%%%%%%%%%%%%%%%%%%%%%%%%%%%%%%%%%%%%

\subsection{$\hat \Cm_{0(0,0)}\times\hat \Bm_1$}

Our final mixed correlation function is between a stress-tensor multiplet and the flavor current. In this case, \eqref{3ptf} reads, 
\begin{align}
 \langle \Jm(z_1)\,\Lm_{ij}(z_2)\,\Om^\Im(z_3) \rangle = \lambda_{\Jm\Lm\Om} \frac{{\hat u}_{i}^{\; k}(z_{23}){\hat u}_{j}^{\; l}(z_{23})}{x_{\bar 3\, 1}^2 x_{\bar 2\,3}^2 x_{\bar 2\, 3}^2 x_{2\,\bar 3}^2 }H_{kl}^I(\Zbf)\,, \label{jlo}
\end{align}
where the $\hat u$ matrices where defined in \eqref{hatu}.

We already saw the implications of the conservation equations of $\Jm$ \eqref{eomJ} and $\Lm$ \eqref{eomL} when we studied the $\Em_q\times \hat\Cm_{0(0,0)}$ and $\Em_q\times\hat\Bm_1$ OPE. The change of position of $\Jm$ from the second point to the first point only interchanges the $\Qm$s for $\Dm$s,
\begin{subequations}
\label{Jcns}
\begin{align}
D^{ij}_1 \langle \Jm (z_1)\,\Lm_{ij}(z_2)\, \Om^\Im(z_3) \rangle=&0 \qquad \Rightarrow\qquad  \bar\Dm^k_{\ad} \bar\Dm^{\ad\, l}H^\Im_{ij}(\Zbf_3)=0\, , \label{Jcns1}\\
\bar D^{ij}_1 \langle \Jm (z_1)\,\Lm_{ij}(z_2)\, \Om^\Im(z_3) \rangle=&0 \qquad \Rightarrow\qquad \Dm^k_\a \Dm^{\a\, l}H^\Im_{ij}(\Zbf_3)=0\, ,\label{Jcns2}
\end{align}
\end{subequations}

The \eqref{Jcns1} condition constraints $H_{kl}(\Zbf)$ to be of the form \cite{Kuzenko:1999pi,Liendo:2015ofa},
\begin{align}
 H_{ij}^\Im(\Zbf)=f(\Xbf,\Th)_{ij}^\Im+g(\Xbf,\Th)_{ijk,\ad}^\Im\Thb^{\ad\, k}+h(\Xbf,\Th)_{ij, \ad\bd}^\Im\Thb^{\ad\bd}\, . \label{sol1}
\end{align} 
The next step is to find the $f$, $g$ and $h$ functions. Since \eqref{Lcns2} does not mix the $\Xbf$ with the $\Th$, it is natural to use this equation to construct the $f$, $g$ and $h$ terms as a $\Th$ expansion,
\begin{align}
 f(\Xbf,\Th)_{(ij)}^{\Im}=\sum_{k=0}^4 f_k(\Xbf)^\Im_{(ij),m_1\cdots m_k,\a_1\cdots \a_k}\Th^{\a_1 m_1}\cdots \Th^{\a_k m_k}\,, \label{CxB:exp}
\end{align}
and similar for $g$ and $h$. The following are the only terms satisfying \eqref{Lcns2},
\begin{align}\begin{array}{llll}
 \epsilon^{(m|(i}\epsilon^{j)|n)}\,, & \e^{m(i}\e^{j)a}\,, & \e^{(m|(i}\e^{j)|n}\e^{o)a}\,, & \Th^{\a\,(m}\e^{n)\,(i}\e^{j)\,a}-\Th^{\a\,a}\e^{(m\,|(i}\e^{j)|\,n)}\,, \\
 \Th^{\a (i}\e^{j)m}\,, &\Th^{\a(i}\e^{j)a}\,, & \Th^{(i\, j)}\,,& \Th^{(i\,j)}\e^{m\,a}\,.
\end{array}\label{CxB:thetasol}\end{align}
Note that \eqref{CxB:thetasol} contains the three structures in \eqref{ExB:thetasol} plus five additional structures. The structures in \eqref{CxB:thetasol} not only tell us the $SU(2)_R$-charge of the $\Om$ operator in \eqref{jlo}, but they also fix its $U(2)_r$-charge thanks to the scaling condition \eqref{scaling}. The final step is to find the suitable functions of $\Xbf$ in \eqref{CxB:exp} satisfying both \eqref{Lcns1} and \eqref{Jcns2}. The physical solutions are,
\begingroup
\allowdisplaybreaks
\begin{subequations}\label{CxB:s}
\begin{align}
 \text{Multiplet} \qquad & H\left(\Zbf \right) \nn \\
  \Am^\Delta_{0,0\left(\frac{\ell}{2},\frac{\ell}{2} \right)}: \qquad & -\frac{1}{2}\left(4+\ell-\Delta \right) \Xbf_{(\a_1(\ad_1}\cdots \Xbf_{\a_{\ell})\ad_{\ell})} \Th^{\m(i}\Xbf_{\m\,\md}\Thb^{\md\,j)}\left(\Xbf^2 \right)^{-3+\frac{\Delta-\ell}{2}} \nn\\
                                                                      &+{\rm i}\left(2-\ell-\Delta \right) \Xbf_{(\a_1(\ad_1}\cdots \Xbf_{\a_{\ell-1}\ad_{\ell-1}}\e_{\ad_\ell)\md}\Xbf_{\a_\ell)\nd}\Thb^{\md\,\nd}\Th^{ij}\left(\Xbf^2 \right)^{-3+\frac{\Delta-\ell}{2}}\nn\\
                                                                      &+\left(\Delta-2 \right) \Xbf_{(\a_1(\ad_1}\cdots \Xbf_{\a_{\ell-1}\ad_{\ell-1}}\Th^{(i}_{\a_\ell)}\Thb^{j)}_{\ad_\ell)}\left(\Xbf^2 \right)^{-2+\frac{\Delta-\ell}{2}} \,, \label{CxB:l1} \\
 \Am^\Delta_{0,0\left(\frac{\ell}{2},\frac{\ell+2}{2} \right)}: \qquad & \Xbf_{(\a_1(\ad_1}\cdots \Xbf_{\a_{\ell})\ad_{\ell}}\left(\frac{{\rm i}}{2}\left(2-\ell-\Delta \right) \Thb_{\ad_{\ell+1}\ad_{\ell+1})}\Th^{ij}\right. \nn \\ &\left.+\Th^{\m(i}\Xbf_{\m(\ad_{\ell+1}}\Thb^{j)}_{\ad_{\ell+2})}\right)\left(\Xbf^2 \right)^{-3+\frac{\Delta-\ell}{2}}\,, \label{CxB:l2}\\
 \Am^\Delta_{0,0\left(\frac{\ell+2}{2},\frac{\ell}{2} \right)}: \qquad & \Xbf_{(\a_1(\ad_1}\cdots \Xbf_{\a_{\ell})\ad_{\ell}}\left(\left(6+\ell-\Delta \right) \Xbf_{\a_{\ell+1}\md}\Xbf_{\a_{\ell+2}\nd}\Thb^{\md\,\nd}\Th^{ij}\right. \nn \\ &\left.-2{\rm i}\Th^{(i}_{\a_{\ell+1}}\Xbf_{\a_{\ell+2})\md}\Thb^{\md\,j)}\right)\left(\Xbf^2 \right)^{-4+\frac{\Delta-\ell}{2}}\,, \label{CxB:l3}\\
 \Cm_{0,0(0,1)}: \qquad & \Th^{\m(i}\Xbf_{\m(\ad_1}\Thb^{j)}_{\ad_2)}\left( \Xbfb^2\right)^{-2}\,, \label{CxB:c0001}\\
 \bar \Cm_{\frac{1}{2},\frac{3}{2}\left(\frac{1}{2},0 \right)}: \qquad & \Xbf_{\a\,\ad}\Thb^{\ad(i}\e^{j)m}\left(\Xbf^2 \right)^{-2}-4{\rm i} \Xbf_{\a\,\ad} \Xbf_{\b\,\bd}\Thb^{\ad\,\bd} \Th^{\b(i}\e^{j)m}\left(\Xbf^2 \right)^{-3}\,, \\
 \Cm_{\frac{1}{2},-\frac{3}{2}\left(0,\frac{1}{2} \right)}: \qquad & \Th^{\m (i}\Xbf_{\m\,\ad_1} \e^{j)m}\left(\Xbf^2 \right)^{-2}\,, \label{CxB:c2}\\
 \Cm_{\frac{1}{2},\frac{3}{2}\left(0,\frac{1}{2} \right)}: \qquad & \Thb_{\ad}^{(i}\e^{j)m}\,, \label{CxB:c3}\\
 \Cm_{\frac{1}{2},\frac{3}{2}\left(\frac{\ell}{2},\frac{\ell+1}{2} \right)}: \qquad &-2 {\rm i} \ell \Xbf_{(\a_1(\ad_1}\cdots \Xbf_{\a_{\ell-1}\ad_{\ell-1}}\Thb_{\ad_{\ell}\ad_{\ell+1})}\Th^{(i}_{\a_{\ell})}\e^{j)m} \nn\\ &+ \Xbf_{(\a_1(\ad_1}\cdots \Xbf_{\a_{\ell}\ad_{\ell}}\Thb_{\ad_{\ell+1}}^{(i}\e^{j)m}\left(\Xbf^2 \right)\,, \label{CxB:c4}\\
 \bar \Cm_{\frac{1}{2},-\frac{3}{2}\left(\frac{\ell+1}{2},\frac{\ell}{2} \right)}: \qquad &\Xbf_{(\a_1(\ad_1}\cdots \Xbf_{\a_{\ell}\ad_{\ell})}\Th^{(i}_{\a_{\ell+1})}\e^{j)m}\,, \\
 \hat\Cm_{1\left(0,0\right)}: \qquad & \e^{(m|(i}\e^{j)|n)}\,, \label{CxB:c5}\\
\nn \hat\Cm_{1\left(\frac{\ell}{2},\frac{\ell}{2}\right)}: \qquad & \Xbf_{(\a_1(\ad_1}\cdots \Xbf_{\a_{\ell-1}\ad_{\ell-1}} \left(\Xbf_{\a_\ell)\ad_\ell)}\e^{(m|(i}\right.\\ &\left.-4 {\rm i}\ell\left(\Th_{\a_\ell}^{(m}\Thb_{\ad_\ell)}^{|(i}+\Th_{\a_\ell}^{a}\Thb_{\ad_\ell)a}\e^{(m|(i} \right) \right)\e^{j)|n)}\,, \label{CxB:c6}\\
 \hat \Bm_{1}: \qquad & -4{\rm i} \Xbf_{\m\,\md}\left(\Th^{\m (n}\Thb^{\md|(i} + \Th^{\m \, a}\Thb^{\md}_a \e^{(n|(i} \right)\e^{j)|m)}\left(\Xbf^2 \right)^{-2}+\e^{(n|(i}\e^{j)|m)}\left(\Xbf^2 \right)^{-1}\,.
\end{align}
\end{subequations}
\endgroup
The discarded solutions to \eqref{jlo} are listed in \eqref{CxB:nu}, \eqref{CxB:nu1} and \eqref{ch1}

The solution \eqref{CxB:c0001} is exactly \eqref{CxB:l2} when it hits its unitarity bound, $\Delta_{UB}=2$. It is also the only physical solution between a family of unphysical ones \eqref{CxB:nu00ll+2}. \eqref{CxB:c2} is also the only physical solution of a larger family \eqref{CxB:unc2}. \eqref{CxB:c4} is valid only for $\ell\ge1$, the case $\ell=0$ being \eqref{CxB:c3}. A similar situation happens with \eqref{CxB:c6}: it is only valid for $\ell\ge1$, the special case $\ell=0$ reduces to \eqref{CxB:c5}, which is discarded.

Unlike the previous cases, the decomposition at the unitarity bound of the long multiplets in \eqref{CxB:s} are unique,
\begin{align}
 \Am^{2+\ell}_{0,0\left(\frac{\ell}{2},\frac{\ell}{2} \right)}\sim &\hat \Cm_{0\left(\frac{\ell}{2},\frac{\ell}{2} \right)}+\hat \Cm_{\frac{1}{2}\left( \frac{\ell-1}{2} , \frac{\ell}{2} \right)} + \hat \Cm_{\frac{1}{2}\left( \frac{\ell}{2} , \frac{\ell-1}{2} \right)}+ \hat \Cm_{1\left( \frac{\ell-1}{2} , \frac{\ell-1}{2} \right)}\,, \label{CxB:long1}\\
 \Am^{2+\ell}_{0,0\left(\frac{\ell}{2},\frac{\ell+2}{2} \right)}\sim &\Cm_{0,0\left(\frac{\ell}{2},\frac{\ell+2}{2} \right)}+ \Cm_{\frac{1}{2},\frac{1}{2}\left(\frac{\ell-1}{2},\frac{\ell+2}{2} \right)}\,, \label{CxB:ub2} \\
 \Am^{2+\ell}_{0,0\left(\frac{\ell+2}{2},\frac{\ell}{2} \right)}\sim &\bar\Cm_{0,0\left(\frac{\ell+2}{2},\frac{\ell}{2} \right)}+\bar\Cm_{\frac{1}{2},-\frac{1}{2}\left(\frac{\ell+2}{2},\frac{\ell-1}{2} \right)}\,.\label{CxB:ub1}
\end{align}
Since we do not find any $\hat \Cm_{\frac{1}{2}\left( \frac{\ell-1}{2} , \frac{\ell+2}{2} \right)}$, $\hat \Cm_{\frac{1}{2}\left( \frac{\ell+2}{2} , \frac{\ell-1}{2} \right)}$, $ \Cm_{\frac{1}{2},\frac{1}{2}\left(\frac{\ell-1}{2},\frac{\ell+2}{2} \right)}$ nor $\bar\Cm_{-\frac{1}{2},-\frac{1}{2}\left(\frac{\ell+2}{2},\frac{\ell-1}{2} \right)}$ solutions, we discard them from the OPE. Note that the decomposition of \eqref{CxB:l1}, \eqref{CxB:long1}, contains the (\ref{CxB:c5},\ref{CxB:c6}) solution, as discussed earlier.

Finally, we write down the $\hat\Cm_{0(0,0)}\times \hat\Bm_1$ OPE,
\begin{align}\label{CxB:OPE} \begin{split}
 \hat\Cm_{0(0,0)}\times \hat\Bm_1 \sim& \Cm_{0,0\left(\frac{\ell}{2},\frac{\ell+2}{2} \right)} + \hat\Cm_{0\left(\frac{\ell+1}{2},\frac{\ell+1}{2} \right)} + \hat \Cm_{1\left(\frac{\ell}{2},\frac{\ell}{2} \right)} + \Cm_{\frac{1}{2},\frac{3}{2}\left(\frac{\ell}{2},\frac{\ell+1}{2} \right)} +\Cm_{\frac{1}{2},-\frac{3}{2}\left(0,\frac{1}{2} \right)} + \hat \Bm_{1}\\
                                      & +\Am^{\Delta}_{0,0\left(\frac{\ell}{2},\frac{\ell+2}{2} \right)}+\Am^{\Delta}_{0,0\left(\frac{\ell}{2},\frac{\ell}{2} \right)}\,, \end{split}
\end{align}
with $\ell\ge0$. Since this OPE is real, we do not write the conjugate multiplets.

\section{Discussion}

Using only $\Nm=2$ superconformal symmetry and Minkowski superspace techniques, we have computed the mixed OPE between a chiral multiplet $\Em_q$, a stress-tensor multiplet $\hat\Cm_{0(0,0)}$, and a flavor current multiplet $\hat\Bm_1$. Those mixed OPEs were obtained by analyzing all possible three-point functions between two of the superfields corresponding to the multiplet listed before and an arbitrary third operator. The solutions were categorized as physical, which we listed in the OPEs \eqref{ExC:OPE}, \eqref{ExB:OPE} and \eqref{CxB:OPE}, and extra solutions listed in the Appendix B. The mixed OPEs involving an $\Em_q$ multiplet have an explicit dependence on its $U(1)_r$-charge. This is not an unexpected result. Computation of two (anti-)chiral multiplets with different $U(1)_r$-charge also shows this behavior \cite{Lemos:2015awa}.

Our results are in complete agreement with the $\hat \Cm_{0(0,0)}\times \hat \Cm_{0(0,0)}$, $\Em_q\times \bar \Em_{-q}$ and $\hat\Bm_1\times \hat\Bm_1$ OPEs previously found. Our mixed OPEs $\Em_q \times \hat\Cm_{0(0,0)}$ and $\hat\Cm_{0(0,0)}\times\hat\Bm_1$ do not contain any $\hat\Cm_{0(0,0)}$ multiplet. This is in agreement with the $\hat \Cm_{0(0,0)}\times \hat \Cm_{0(0,0)}$ OPE \cite{Liendo:2015ofa}, which does not contain any $\hat\Bm_1$ nor $\Em_q$ operators. From the OPE between a chiral and an anti-chiral multiplet \cite{Dolan:2004mu}, 
%\begin{align}
% \Em_q\times\bar\Em_{-q}\sim \Im+\hat \Cm_{0\left(\frac{\ell}{2},\frac{\ell}{2} \right)}+\Am^\Delta_{0,0\left(\frac{\ell}{2},\frac{\ell}{2} \right)}\,,
%\end{align}
it was expected to obtain a chiral multiplet $\Em_q$ from the $\Em_q\times \hat\Cm_{0(0,0)}$ OPE, and neither a $\hat\Bm_1$ nor $\Em_q$ in the $\Em_q\times\hat\Bm_1$ OPE. Finally, our $\hat \Cm_{0(0,0)}\times\hat\Bm_1$ OPE contains a $\hat\Bm_1$ multiplet in the expansion, which agrees with the $\hat\Bm_1\times \hat\Bm_1$ OPE \cite{Fitzpatrick:2014oza}.

An interesting generalization of this work is to find the OPEs between different $\hat\Bm_R$ multiplet, with $R>1$ and the $\hat\Cm_{0(0,0)}$ multiplet
%, and hopefully and analytical form for any $R$
. As mentioned early, bounds for the central charge and the flavor central charge were obtained using the,
\begin{align}
 \langle \hat \Cm_{0(0,0)} \hat \Cm_{0(0,0)} \hat \Cm_{0(0,0)} \hat \Cm_{0(0,0)} \rangle\,, \qquad\qquad \langle \hat \Bm_1 \hat \Bm_1 \hat \Bm_1 \hat \Bm_1 \rangle \,\qquad \text{and} \qquad \langle \hat \Cm_{0(0,0)} \hat \Cm_{0(0,0)} \hat \Bm_1 \hat \Bm_1 \rangle \,,
\end{align}
correlators and the chiral algebra correspondence \cite{Beem:2013sza,Liendo:2015ofa,Lemos:2015orc}. When those bounds are saturated, the OPE coefficients of certain allowed operators also vanish. For example, the bound that comes from the stress-tensor four-point function, $c\ge\frac{11}{30}$, is saturated only if the OPE coefficient of the $\hat \Cm_{1(0,0)}$ multiplet that appear in the $\hat \Cm_{0(0,0)}\times \hat \Cm_{0(0,0)}$ is 0. The theory with $c=\frac{11}{30}$ corresponds to the simplest Argyres-Douglas fixed point $H_0$. Using the superconformal index, it was confirmed that this multiplet does not appear in the $\hat \Cm_{0(0,0)}\times \hat \Cm_{0(0,0)}$ OPE in the $H_0$ theory \cite{Song:2015wta}. By studying the OPEs involving a $\hat \Bm_R$ multiplet with higher $SU(2)_R$-charge, the chiral algebra correspondence should yield stronger bounds for the theory. Furthermore, the saturation of the bounds will imply the vanishing of certain OPE coefficient, as in the $H_0$ case. This vanishing of OPE coefficients can be given as input in the numerical bootstrap in order to single out a particular theory.

%Finally, the next natural and necessary step before we can apply the bootstrap program is to write down the superconformal block expansions. 

%There are several approaches to compute the conformal and superconformal blocks available in the literature with different degrees of success \cite{Hadasz:2006qb,Hadasz:2007ns,Suchanek:2010zz,Costa:2011mg,Dolan:2011dv,Costa:2011dw,Belavin:2012uf,Hadasz:2012im,Hanany:2012mb,SimmonsDuffin:2012uy,Dymarsky:2013wla,Hogervorst:2013kva,Fitzpatrick:2014oza,Khandker:2014mpa,Elkhidir:2014woa,Costa:2014rya,Bissi:2015qoa,Rejon-Barrera:2015bpa,Doobary:2015gia,Echeverri:2015rwa}. Usually this is a very long a cumbersome computation and we will leave it for a next article. 

\section*{Acknowledgments}

I want to thank P. Liendo, who not only helped me to start this project, but also encouraged me to complete it and gave me useful advise, comments and suggestions. Additionally I want to thank S. Retamales, for her comments and much more. I also want to thank the hospitality of both Humboldt-Universit\"at zu Berlin, where this project started, and Universidad Nacional Andr\'es Bello, where this project was completed. The author was supported by CONICYT project No. 21120105.
%I want to thank my mom for carrying me during nine months. Thanks Mom!

\appendix

\section{Long, short and semi-short multiplets}
Representation theory of the $\Nm=2$ superconformal algebra has been extensively used during this work. We follow the notation of \cite{Dolan:2002zh}, where all unitary irreducible representations of the extended superconformal algebra were constructed. The $\Nm=2$ superconformal algebra $SU(2,2|2)$ contains as a subalgebra the conformal algebra $SU(2,2)$ generated by $\{\Pm_{\a\,\ad},\Km_{\a\,\ad},\Mm^\a_{\;\b},\bar \Mm^{\ad}_{\;\bd},\Dm \}$, where $\a=\pm$ and $\ad=\dot\pm$ are the Lorentz indices. $SU(2,2|2)$ also contains an $R$-symmetry algebra $SU(2)_R\times U(1)_r$ with generators $\{R^i_{\;j},r\}$, where the $i,j=1,2$ are the $SU(2)_R$ indices. Along with the bosonic charges, there are fermionic supercharges, the Poincar\'e and conformal supercharges, $\{\Qm_\a^i,\bar\Qm_{\ad\,i},\Sm^\a_i,\bar\Sm^{\ad\,i}\}$. 

The spectrum of operators of $SU(2,2|2)$ is constructed from its highest weight, or superconformal primary. Acting with the Poincar\'e supercharges on the superconformal primary, superconformal descendants are generated. A general superconformal primary is denoted by $\Am^\Delta_{R,r(j,\bar j)}$, and is referred to as long multiplet. The only restriction for such multiplet is to satisfy a unitary bound \cite{Dobrev:1985qv},
\begin{align}
 \Delta\ge 2+2j+2R+r\,,\,2+2\bar j+2R-r\,.
\end{align}
If the multiplet is annihilated by a certain combination of $\{\Qm_\a^i,\bar\Qm_{\ad\,i}\}$ is referred to as short or semi-short and it saturates the unitarity bound. These combinations are,
\begin{align}
 \Bm^i:\qquad & \Qm^i_{\a}\Psi=0\,,\\
 \bar \Bm^i:\qquad & \bar\Qm^i_{\ad}\Psi=0\,,\\
 \Cm^i:\qquad &\left\lbrace \begin{array}{ll} \e^{\a\b}\Qm^i_{\a} \Psi_{\b} & j\neq 0\,, \\ \e^{\a\b}\Qm^i_{\a}\Qm^i_{\b} \Psi & j=0\,, \end{array} \right.\\
 \bar \Cm^i:\qquad &\left\lbrace \begin{array}{ll} \e^{\ad\bd}\bar \Qm^i_{\ad} \Psi_{\bd} & \bar j\neq 0\,, \\ \e^{\ad\bd}\bar \Qm^i_{\ad}\bar \Qm^i_{\bd} \Psi & \bar j=0\,. \end{array} \right.
\end{align}
$\Bm$-type conditions are called short while $\Cm$-type are called semi-short, because the former are stronger conditions. In Tab. \ref{repre} we list all possible shortening conditions.

\begin{table}
\centering
\begin{tabular}{|l|l|l|l|}\hline
Shortening & \multicolumn{2}{|c|}{Unitarity bounds} & Multiplet\\ \hline
 & $\Delta>2+2j+2R+r$ & $\Delta>2+2\bar j +2R-r$ & $\Am^\Delta_{R,r\left(j,\bar j \right)}$\\ \hline
$\Bm^1$ & $\Delta=2R+r$ & $j=0$ & $\Bm_{R,r(0,\bar j)}$\\ \hline 
$\bar \Bm_2$ & $\Delta=2R-r$ & $\bar j=0$ & $\bar \Bm_{R,r(j,0)}$\\ \hline 
$\Bm^1 \cap \Bm^2$ & $\Delta=r$ & $R=\bar j=0$ & $\Em_{r(0,\bar j)}$\\ \hline 
$\bar \Bm_1 \cap \bar \Bm_2$ & $\Delta=-r$ & $R=j=0$ & $\bar \Em_{r(j,0)}$\\ \hline 
$\Bm^1\cap \bar \Bm_2$ & $\Delta=2R$ & $j=\bar j=r=0$ & $\hat \Bm_{R}$\\ \hline
$\Cm^1$ & $\Delta=2+2j+2R+r$ & & $\Cm_{R,r(j,\bar j)}$\\ \hline 
$\bar \Cm_2$ & $\Delta=2+2\bar j+2R-r$ & & $\bar \Cm_{R,r(j,\bar j)}$\\ \hline 
$\Cm^1 \cap \Cm^2$ & $\Delta=2+2j+r$ & $R=0$ & $\Cm_{0,r(j,\bar j)}$\\ \hline 
$\bar \Cm_1 \cap \bar \Cm_2$ & $\Delta=2+2\bar j-r$ & $R=0$ & $\bar \Cm_{0,r(j,\bar j)}$\\ \hline 
$\Cm^1\cap \bar \Cm_2$ & $\Delta=2+j+\bar j+2R$ & $r=\bar j-j$ & $\hat \Cm_{R(j,\bar j)}$\\ \hline
$\Bm^1 \cap \bar\Cm_2$ & $\Delta=1+\bar j+2R$ & $r=\bar j+1$ & $\Dm_{R(0,\bar j)}$\\ \hline 
$\bar \Bm_2 \cap \Cm^1$ & $\Delta=2+j+2R$ & $-r=j+1$ & $\bar \Dm_{R(j,0)}$\\ \hline 
$\Bm^1\cap \Bm^2 \cap \bar\Cm_2$ & $\Delta=r=1+\bar j$ & $R=0$ & $\Dm_{0(0,\bar j)}$\\ \hline 
$\bar \Bm_1\cap \bar \Bm_2 \cap \Cm^1$ & $\Delta=-r=1+j$ & $R=0$ & $\bar \Dm_{0(j,0)}$\\ \hline 
\end{tabular}
\label{repre}\caption{All possible short and semi-short representations for the $\Nm=2$ superconformal algebra.}
\end{table}

The decomposition of a long multiplet $\Am^\Delta_{R,r\left(j,\bar j \right)}$ when it hits its unitarity bound is given by,
\begin{align}\label{decom}
 \begin{split}
  \Am^{2+2j+2R+r}_{R,r\left(j,\bar j \right)}\sim& \Cm_{R,r\left(j,\bar j \right)} + \Cm_{R+\frac{1}{2},r+\frac{1}{2}\left(j-\frac{1}{2},\bar j \right)}\,,\\
  \Am^{2+2j+2\bar j+2R}_{R,r\left(j,\bar j \right)}\sim& \hat \Cm_{R\left(j,\bar j \right)} +\hat \Cm_{R+\frac{1}{2}\left(j-\frac{1}{2},\bar j \right)}+\hat\Cm_{R+\frac{1}{2}\left( j,\bar j-\frac{1}{2}\right)}+\hat \Cm_{R+1\left(j-\frac{1}{2},\bar j-\frac{1}{2} \right)}\,,\\
  \Am^{2+2\bar j+2R-r}_{R,r\left(j,\bar j \right)}\sim& \bar \Cm_{R,r\left(j,\bar j \right)} + \bar \Cm_{R+\frac{1}{2},r-\frac{1}{2}\left(j,\bar j -\frac{1}{2}\right)}\,.
 \end{split}
\end{align}

\section{Discarded solutions}

Several solutions to \eqref{pjo}, \eqref{plo} and \eqref{jlo} were not listed in the corresponding OPE, because we regarded them as unphysical. We categorize them in three types,
\begin{itemize}
\item Non-unitary solutions. Those solutions have a conformal dimension below the unitarity bound corresponding to their quantum numbers.
\item Long multiplets with fixed dimension. Long multiplets with fixed dimensions were argued to come from a theory with extended $\Nm=4$ symmetry \cite{Liendo:2015ofa}. We are only interested in theories with $\Nm=2$, thus, we will consider such multiplets as being irrelevant to $\Nm=2$ dynamics.
\item Solutions with a vanishing overall coefficient. We also find a case where the solution to the three-point function corresponds to a physical multiplet, a stress-tensor multiplet. Uniqueness of the stress-tensor implies another symmetry of the three-point function, which is only satisfied if the overall coefficient vanishes.
\end{itemize}

\subsubsection*{$\Em_q\times \hat \Cm_{0(0,0)}$}
There are two types of discarded solutions to \eqref{pjo} which are not listed in \eqref{ExC:OPE}: non-unitary and solution corresponding to a long multiplet with fixed dimension. The non-unitary solutions are,
\begin{subequations}\label{ExC:nu}
\begin{align}
\left(\Delta,R,r,j,\bar j \right) \qquad  & H\left(\Zbf \right) \nn \\
\left(\frac{1}{2}+q, \frac{1}{2}, \frac{3}{2}-q, 0,\frac{1}{2} \right):\qquad   & \left(\Xbfb^2 \right)^{-1}\Thb^i_{\ad}\,,\\
\left(q-\ell,0,-q,\frac{\ell}{2},\frac{\ell}{2} \right):\qquad &  \left(\Xbfb^2 \right)^{-1-\ell}\,\Xbfb_{\a_1\,\ad_1}\cdots \Xbfb_{\a_{\ell}\,\ad_{\ell}} \\
\left(-\frac{1}{2}+q-\ell,\frac{1}{2},\frac{3}{2}-q,\frac{\ell+1}{2},\frac{\ell}{2} \right): \qquad & \left(\Xbfb^2 \right)^{-2-\ell}\Xbfb_{\a_1\,\ad_1}\cdots \Xbfb_{\a_{\ell}\,\ad_{\ell}}\Xbfb_{\ad_{\ell+1}\,\md}\Thb^{\md\,i}\,.
\end{align}
\end{subequations}
Although it is puzzling to find solutions with a conformal dimension that decreases with the spin, this kind of solutions are not new. They have already appeared in $\Nm=1$ theories when computing the three-point function with two flavor currents \cite{Berkooz:2014yda} and in $\Nm=2$ theories when studying the three-point function with two stress-tensor multiplets \cite{Liendo:2015ofa}.

The only long multiplet with fixed dimension is,
\begin{align}\label{ExC:n4}
\Am^{\frac{7}{2}+q+\ell}_{\frac{1}{2},\frac{3}{2}-q\left(\frac{\ell+1}{2},\frac{\ell}{2}\right)}: \qquad& \Xbfb_{\a_1\,\ad_1}\cdots \Xbfb_{\a_{\ell}\,\ad_{\ell}}\Xbfb_{\ad_{\ell+1}\,\md}\Thb^{\md\,i}\,.
\end{align}

\subsubsection*{$\Em_q\times \hat \Bm_1$}

The only unphysical solution to \eqref{plo} is,
\begin{align}
 \left(-\frac{3}{2}+q-\ell,\frac{1}{2},\frac{3}{2}-q,\frac{\ell+1}{2},\frac{\ell}{2} \right): \qquad& \left(\Xbfb^2 \right)^{-2-\ell}\Xbfb_{\a_1\,\ad_1}\cdots \Xbfb_{\a_\ell\,\ad_\ell}\Xbfb_{\a_{\ell+1}\,\bd}\Thb^{\bd(i}\e^{j)m}\,. \label{ExB:nu}
\end{align}
Since the conformal dimension of \eqref{ExB:nu} is below the unitarity bound for its quantum numbers, $\Delta_{UB}=\frac{11}{2}-q+\ell$ for $4\ge q \ge1$ and $\Delta_{UB}=\frac{3}{2}+q+\ell$ for $q\ge4$, we regard it as a non-unitarity solution.

\subsubsection*{$\hat \Cm_{0(0,0)} \times\hat\Bm_1$}

The non-unitary solutions to \eqref{jlo} are,
\begingroup
\allowdisplaybreaks
\begin{subequations}\label{CxB:nu}
\begin{align}
 \left(\Delta,R,r,j,\bar j \right) \qquad & H\left(\Zbf \right) \nn \\
 \left(\frac{3}{2}-\ell,\frac{1}{2},-\frac{3}{2},\frac{\ell+1}{2},\frac{\ell}{2}\right): \qquad &\Xbf_{(\a_1(\ad_1}\cdots\Xbf_{\a_{\ell-1}\ad_{\ell-1}}\left(\Xbf_{\a_{\ell}\ad_{\ell})}\Xbf_{\a_{\ell+1}\bd}\Thb^{\bd (i}\e^{j)m}\left(\Xbf^2 \right)^{-2-\ell} \right.\nn \\
                                                                                               & -2{\rm i}\left(2+\ell\right) \Xbf_{\a_{\ell}\ad_{\ell})}\Xbf_{\a_{\ell+1})\bd}\Xbf_{\m\,\md}\Thb^{\bd\,\md}\Th^{\m(i}\e^{j)m}\left(\Xbf^2 \right)^{-3-\ell} \nn\\
                                                                                               &\left. -2{\rm i} \ell\e_{\ad_{\ell}\bd}\Xbf_{\a_{\ell})\,\md}\Thb^{\md\,\bd} \Th^{(i}_{\a_{\ell+1})} \e^{j)m}\left(\Xbf^2 \right)^{-2-\ell}\right) \,, \label{CxB:unc2}\\
 \left(1-\ell,0,0,\frac{\ell}{2},\frac{\ell}{2}\right): \qquad &  \Xbf_{(\a_1(\ad_1}\cdots\Xbf_{\a_{\ell-1}\ad_{\ell-1}}\left( \ell \Th_{\a_\ell)}^{(i}\Thb_{\ad_\ell)}^{j)}\left(\Xbf^2 \right)^{-1-\ell} \right. \nn\\
                                                               & \left. (1+\ell)\Xbf_{\a_{\ell})\ad_{\ell})} \Th^{\m (i}\Xbf_{\m\,\md}\Thb^{\md \, j)} \left(\Xbf^2 \right)^{-2-\ell} \right) \\
 \left(2,0,0,\frac{\ell}{2},\frac{\ell}{2} \right): \qquad & \Xbf_{(\a_1(\ad_1}\cdots\Xbf_{\a_{\ell-1}\ad_{\ell-1}} \left(\Xbf_{\a_{\ell})\ad_{\ell})} \Th^{\m (i}\Xbf_{\m\,\md}\Thb^{\md \, j)}\right. \nn \\
                                                           & \left.-{\rm i}\ell \Th^{ij}\e_{\ad_\ell)\md}\Xbf_{\a_{\ell}\nd}\Thb^{\md\,\nd}\right)\left(\Xbf^2 \right)^{-(4+\ell)/2}\,, \\
 \left(2-\ell,0,0,\frac{\ell}{2},\frac{\ell+2}{2}\right): \qquad & \Xbf_{(\a_1(\ad_1}\cdots\Xbf_{\a_{\ell})\ad_{\ell})} \Th^{\m (i}\Xbf_{\m\,\ad_{\ell+1}}\Thb^{ j)}_{\ad_{\ell})}\left( \Xbfb^2\right)^{-2-\ell}\,, \label{CxB:nu00ll+2}\\
 \left(2-\ell,1,0,\frac{\ell}{2},\frac{\ell}{2} \right): \qquad & \Xbf_{(\a_1\ad_1}\cdots \Xbf_{\a_{\ell-1}\ad_{\ell-1}}\left( \Xbf_{\a_{\ell})\ad_{\ell})}\e^{(m|(i}\e^{j)|n)}\left(\Xbf^2 \right)\right.\nn \\
                                                                & -4 {\rm i}(\ell+1) \Xbf_{\a_{\ell})\ad_{\ell})} \Xbf_{\m\,\md}\left( \Th^{\m(m}\Thb^{\md\, |(i} +\Th^{\m\,a}\Thb^{\md}_a\e^{(m|(i} \right) \e^{j)|n)}\nn \\
                                                                & \left.-4 \ell \left( \Th^{(m}_{\a_\ell)}\Thb^{|(i}_{\ad_\ell)}+\Th^{a}_{\a_\ell)}\Thb_{\ad_\ell)a}\e^{(m|(i} \right) \e^{j)|n)}\left(\Xbf^2 \right)\right)\left(\Xbf^2 \right)^{-2-\ell}\,, \\
 \left(\frac{3}{2}-\ell,\frac{1}{2},-\frac{3}{2},\frac{\ell}{2},\frac{\ell+1}{2} \right): \qquad & \Xbf_{(\a_1\ad_1}\cdots \Xbf_{\a_{\ell})\ad_{\ell}}\Th^{\m (i}\Xbf_{\m\,\ad_{\ell+1}} \e^{j)m}\left(\Xbf^2 \right)^{-\ell-2}\,.
\end{align}
\end{subequations}
\endgroup

We also find a solution to \eqref{jlo} which corresponds to a long multiplet with fixed dimension,
\begin{align}\label{CxB:nu1}
 \Am^{6+\ell}_{0,0\left(\frac{\ell}{2},\frac{\ell}{2}\right)}: \qquad & \Xbf_{(\a_1(\ad_1}\cdots\Xbf_{\a_{\ell}\ad_{\ell})}\Th_{\a_{\ell+1}}^{(i}\Xbf_{\a_{\ell+2})\md}\Thb^{\md\,j)}\,.
\end{align}
As explained before, we regard this solution as coming from a theory with enhanced $\Nm=4$ symmetry.

Finally, there is a very special solution to \eqref{jlo},
\begin{align}
 H\left(\Zbf \right)=\Th^{\a(i}\Xbf_{\a\ad}\Thb^{\ad\,j)}\left(\Xbf^2 \right)^{-2}\,, \label{ch1}
\end{align}
which has conformal dimension $\Delta=2$. This solution belongs to a $\hat\Cm_{0(0,0)}$ multiplet, corresponding to a stress-tensor multiplet. Studying the $\hat \Cm\times \hat \Cm$ \cite{Kuzenko:1999pi,Liendo:2015ofa} we know that a $\hat\Cm$-multiplet cannot appear in the OPE of $\hat \Cm\times \hat\Bm$. It seems puzzling that we obtain such a solution. But it is already known that this solution, although satisfies (\ref{Lcns}-\ref{Jcns}), it is not symmetric under $z_1\leftrightarrow z_3$, which comes from the uniqueness of the stress-tensor. Thus, the proportionality constant of \eqref{ch1} has to be 0 (see section 3.3.3 of \cite{Kuzenko:1999pi}.) This is the only case where there is another condition besides the constraints that comes from the EOM of the $\Jm$ and $\Lm_{(ij)}$ multiplets that is not satisfied.

\bibliographystyle{JHEP}
\bibliography{bibliography}{}

\end{document}